\documentstyle[12pt,aasms4,psfig]{article}

\journalid{VOL}{JOURNAL DATE}
\articleid{START PAGE}{END PAGE}
\paperid{MANUSCRIPT ID}

\cpright{AAS}{1996}
\ccc{CODE}

\lefthead{Petrosian \& Lee}
\righthead{Gamma-Ray Burst Fluence}

\begin{document}

\title{The Fluence Distribution of Gamma-Ray Bursts}
\author{Vah\'{e} Petrosian\altaffilmark{1}
and Theodore T. Lee\altaffilmark{2}}
\affil{Center for Space Science and Astrophysics,
Stanford University, Stanford, CA 94305-4060}

\altaffiltext{2}{Department of Applied Physics}
\altaffiltext{1}{Departments of Physics and Applied Physics}

\begin{abstract}
In the use and interpretation of 
$\log{N}$--$\log{S}$ distributions for gamma-ray
bursts, burst peak flux has typically been used for $S$.
We consider here the use of the fluence as a measure of $S$, 
which may be a more appropriate quantity than the peak flux
in such highly variable sources.  We demonstrate how using the
BATSE trigger data we can determine the selection effects on fluence.
Then using techniques developed elsewhere to account for the important
threshold effects and correlations.  Applying the 
appropriate corrections to the distributions, we obtain a fluence distribution
which shows a somewhat sharper break than the peak flux distribution,
implying a possibly narrower fluence luminosity distribution.
If bursts are at cosmological distances, these observations together
indicate that evolution of the luminosity function is required.
\end{abstract}

\section{Introduction}

Because of their transient nature and lack of counterparts, 
the distance to Gamma-Ray Bursts (GRBs) is not known.  Statistical
distributions have therefore been used as indirect 
means of estimating their distances.  The original discovery by
the Burst and Transient Source Experiment (BATSE) team (Meegan et al.
1992), and the subsequent confirmations (Meegan 1996) 
that GRBs are isotropically distributed in the sky has 
gradually strengthened cosmological interpretations of 
GRB sources.  
If bursts are indeed of a cosmological origin, 
then the so-called $\log{N}$-$\log{S}$ distribution can be used 
to constrain the model parameters.

Under the assumption that bursts are distributed homogeneously and isotropically
in a static Euclidean space (HISE), 
the logarithmic slope of the $\log{N}$-$\log{S}$
distribution is expected to be -3/2.  In cosmological scenarios, 
the observed deviations from
this slope are due to 
the breakdown of the last two conditions in HISE.
The degree of deviation depends on not only the geometry and expansion rate
of space but also the shape and evolution of the luminosity function.
This latter effect is less significant for 
narrow and slowly evolving luminosity functions, but if
the range of luminosities is larger than the observed range of $S$,
it will obscure the cosmological effects.
Therefore
the choice of the parameter $S$ is important in the interpretation
of the $\log{N}$-$\log{S}$ results.

For steady sources, this choice is obvious;
$S$ would be represented by the steady and well-defined photon or energy flux.
However, for highly transient sources such as GRBs, the measured flux
depends on the time scale over which the burst is observed.  
Ideally, one would like the flux measure to be instantaneous, defined
such that the observational time scale for the accumulation of photons
is smaller than the intrinsic variation time scale of the source.
In practice, observational measures such as the ``peak flux'' have
been averaged over a time interval $\Delta t$ which might not be
small compared to the intrinsic time scale.  Use of this version of
the ``peak flux''
for the parameter $S$ in the $\log{N}$-$\log{S}$ distributions can lead to
ambiguous interpretations, and a number of complicated steps have to be
taken in order to extract the instantaneous peak flux distribution 
from the data (see Lee \& Petrosian 1996a, hereafter LP).

These complications can be 
avoided by using a different observational
measure for $S$.  
One such measure is the
fluence ${\cal F}$, which we define to be the total (time-integrated)
radiant energy per unit
area within the BATSE trigger range of 50--300 keV.
The fluence distribution may prove to be a more useful tool 
in cosmological studies than the peak flux distribution if 
the time-integrated luminosity has
a narrower intrinsic dispersion and undergoes less evolution
than the peak luminosity.
In fact, given the wide dispersion in the durations of GRBs, it is difficult to
justify why the intensity of the highest spike in a bursting source
should have a narrow distribution.
Perhaps the total energy released has a narrower distribution than
that of the peak luminosity.  For example, it is more likely
that the total energy released in a compact
object merger
might be a more appropriate ``standard
candle'' than the rate of energy release in such a merger.
Furthermore, in cosmic fireball models which presumably describe
the physics of these mergers, the relationship between the 
luminosity observed in the detector's rest frame to that emitted by the source
depends strongly on the bulk Lorentz factor of the
expanding shell (Meszaros \& Rees 1993, 1994; 
Madras \& Fenimore 1996), which is not likely
to be universal for all bursts.

In this paper we investigate the $\log{N}$-$\log{{\cal F}}$ 
relation for BATSE GRBs.
In the next section we describe how a bias-free fluence distribution
can be obtained from instruments such as BATSE which trigger on some
average flux value.  In \S \ref{sec-fludiscuss} we discuss our
results.

\section{Analysis}

\subsection{Obtaining the Fluence Limit}

We use the publicly available BATSE 3B catalog data, 
which provides the maximum and
minimum photon counts $C_{max}$ and $C_{min}$ 
for the three trigger time intervals $\Delta
t = 64, 256, 1024$~ms.
The most fundamental selection effect is that
$C_{max}$
must exceed the threshold $C_{min}$.
We are interested in the more physically meaningful fluences and fluxes.
The BATSE catalog also gives values of the total energy fluence ${\cal F}$
within the 50--300 keV range, along with three measures of the average 
peak flux.
The average peak flux is
given by $\bar{f}_P = C_{max}/ (A_{eff}(\theta,\phi)\Delta t)$,
where $A_{eff}$ is the effective detector area (including the spectral response)
of the instrument in the direction $(\theta, \phi)$.  

\begin{figure}
\psfig{file=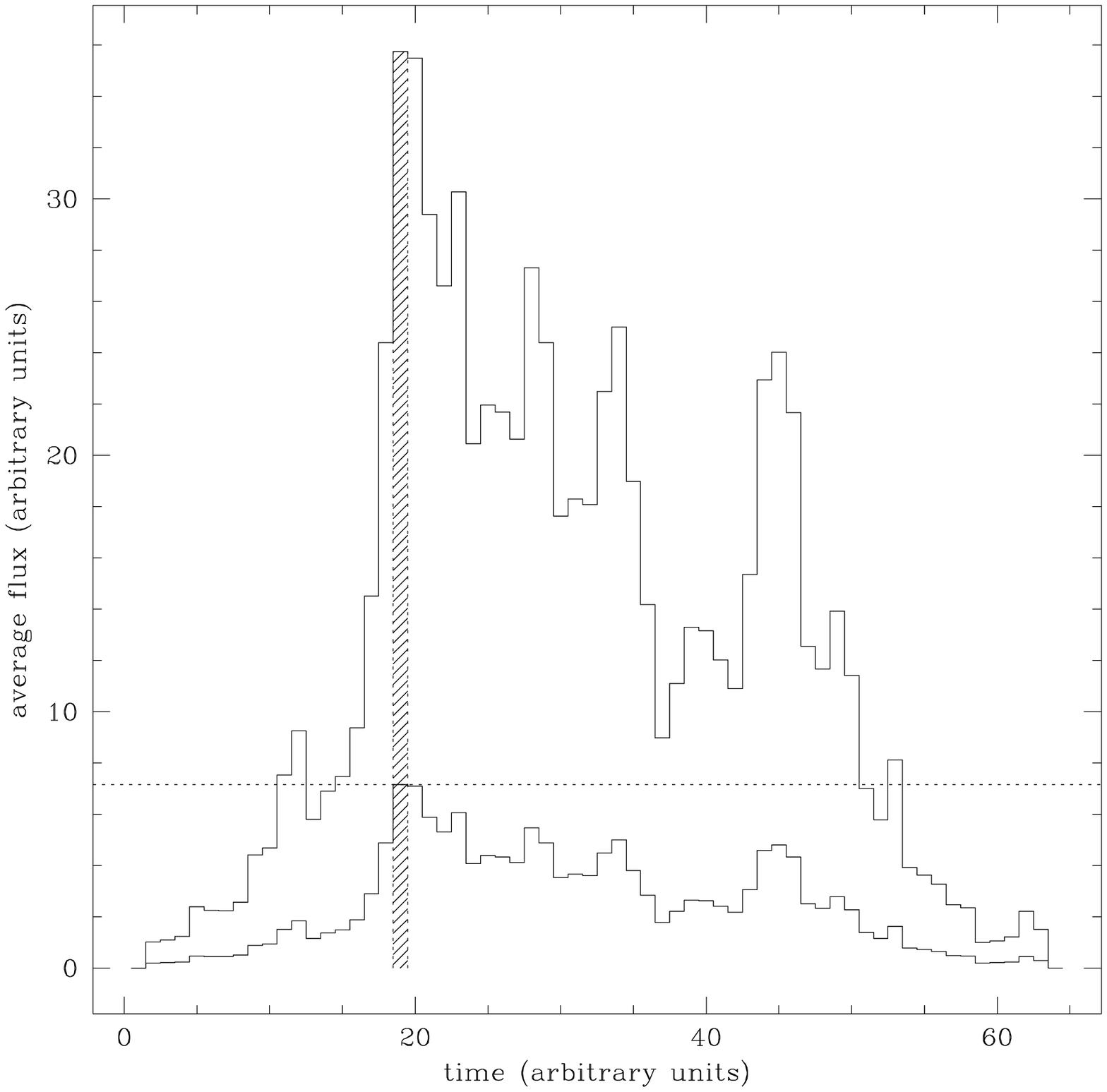,width=5.875in}
\caption{
Schematic diagram of burst selection.  The larger histogram represents 
the light curve of a typical burst.  Neglecting spectral variation effects, the
fluence is proportional to the
time integral of the curve.  The shaded bin represents the
peak flux of the burst (integrated over the trigger time), 
and the dotted line represents the average value of
the limiting flux, below which the burst would not have triggered.
The smaller histogram shows the light curve scaled down by a factor
of $\bar{f}_P/\bar{f}_{lim}$ (or equivalently $C_{max}/C_{min}$). 
No matter what the light curve looks like, 
it can be seen that if $C_{max}/C_{min}$ is constant
throughout the burst, then the limiting fluence for the burst is
simply given by the fluence scaled by $C_{max}/C_{min}$.
}
\label{fig:schematic}
\end{figure}

Given $\bar{f}_P$ or ${\cal F}$, we 
ask what would have been the threshold for detection of a burst
for any of these quantities.  It is easy to see that a burst with
average peak flux $\bar{f}_P$ coming from a direction $\theta$ and $\phi$
would trigger BATSE if $\bar{f}_P > \bar{f}_{lim} \equiv 
C_{min} /(A_{eff}(\theta,\phi)\Delta t)$.  Otherwise $C_{max}$ would
be less than $C_{min}$.  Therefore, the threshold on the average peak
flux is $\bar{f}_{lim} = \bar{f}_P C_{min}/C_{max}$.
The same relation also exists for the fluence and its limit, as can
be seen schematically by examining Figure \ref{fig:schematic}.  
A burst with observed fluence ${\cal F}$ (or $\bar{f}_P$)
and a particular pulse
profile, spectrum, etc. would have been undetected if its fluence (or flux) was lowered by a 
factor of $C_{max}/C_{min}$, because then its peak counts $C_{max}$
would have been
less than the limiting counts $C_{min}$.
Clearly then ${\cal F}_{lim} =
{\cal F} C_{min}/C_{max}$, as long as the background count rate
does not vary significantly throughout the duration of the burst.
In summary,
\begin{equation}
\frac{C_{max}}{C_{min}} = \frac{\bar{f}_P}{\bar{f}_{lim}} 
= \frac{{\cal F}}{{\cal F}_{lim}}. \label{eq:fluflulim}
\end{equation}
The last equality is approximate because 
of the implicit conversion between photon counts and energy.
This will be a good approximation if the source spectra do not
change drastically throughout the duration of the burst.

Note that all of the details of the pulse profile, spectrum, and
instrumental response are hidden in the ratio.  The problem of
obtaining the distribution 
of any of the quantities in equation (\ref{eq:fluflulim}) can
be described generally as the problem of 
obtaining the distribution of some variable
$x$ subject to the condition that $x_i > x_{i,lim}$ for each data
point $i$.  
A general solution to this problem was described by Petrosian (1993)
and has been extensively discussed
in LP.
It can be seen that
extracting the fluence distribution is in principle
no different from extracting
the peak flux distribution.  
It should also be noted that the conclusions drawn from the distribution
of $V/V_{max}$ or its average are unchanged no matter which of these
properties is used as a measure of distance.

\subsection{Fluence Limit Interpretation} \label{sec-interp}

Although the extraction of the fluence distribution is computationally
straightforward, the interpretation
of ${\cal F}_{lim}$ differs from that of $C_{min}$ and hence
deserves some explanation.
$C_{min}$ is simply a threshold that
depends only on the background count rate and is a variable independent
of the physical burst properties.  
For bursts which have durations $T \ll \Delta t$ it is clear that
${\cal F} = C_{max}\langle h\nu \rangle / A_{eff}(\theta,\phi)$ and ${\cal F}_{lim} 
= C_{min}\langle h\nu \rangle / A_{eff}(\theta,\phi)$ are independent, 
where $\langle h\nu \rangle$
is the average photon energy in the 50--300 keV range.
For long duration bursts with $T > \Delta t$, we have approximately
${\cal F} \propto \bar{f}_P T \langle h\nu \rangle$, so that from
equation (\ref{eq:fluflulim}) we obtain
\begin{equation}
{\cal F}_{lim} = \langle h\nu \rangle T \frac{C_{min}}{A_{eff}(\theta,\phi)\Delta t},
\end{equation}
indicating that if as expected $\langle h\nu \rangle$, $A_{eff}(\theta,\phi)$,
and $C_{min}$
are essentially independent of ${\cal F}$, the fluence limit is approximately proportional to the
duration.

\subsection{Fluence Distributions}

We now examine all bursts for which a fluence measurement and a
value of $C_{max}/C_{min}$ exists.  The values of $C_{max}/C_{min}$ are
known for three time scales: 64~ms; 256~ms; and 1024~ms.  We use
the 1024~ms values of $C_{max}/C_{min}$ because this time scale is the
most sensitive and allows us to use the largest number of bursts,
but we also note that the other time scales give essentially
identical results.  
Note that unlike in the determination of the distribution of
instantaneous peak flux (see LP), 
where a correction for the short duration bias based on
some observational duration measure was
necessary, there is no need to have a separate measure of duration 
to determine the fluence limit.
This increases the number of bursts available from 514 to 555 for the 1024~ms
trigger.  It may be argued that bursts with no well-defined durations
may have data gaps or some other problems which would make
the fluence measurements unreliable.  As it turns out, the resulting
fluence distributions are insensitive to whether or not we include the
extra 41 bursts.

\begin{figure}
\psfig{file=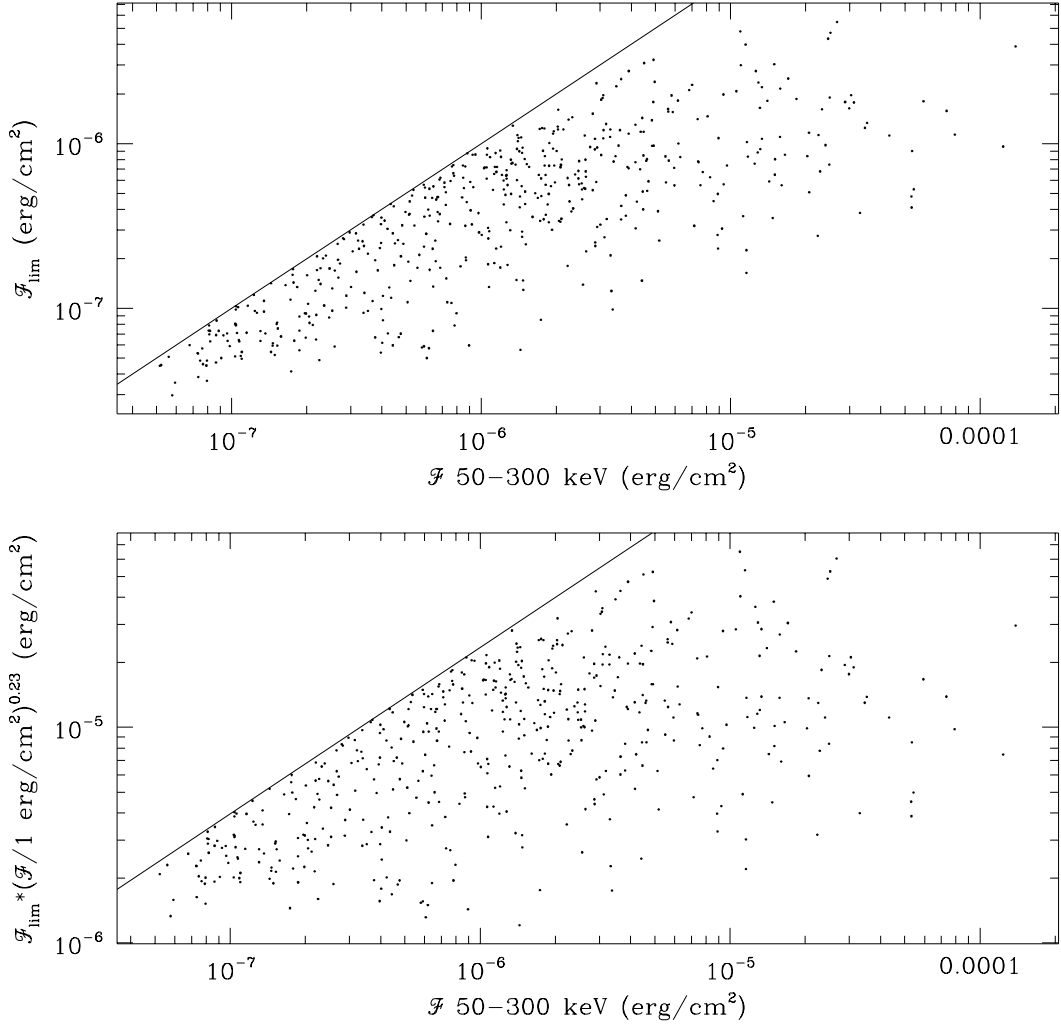,width=5.875in}
\caption{
{\em top panel:} Bivariate distribution of ${\cal F}$ and ${\cal F}_{lim}$.
The diagonal line indicates the selection criterion ${\cal F} > 
{\cal F}_{lim}$.  {\em bottom panel:} Bivariate distribution of ${\cal F}$ 
and ${\cal F}_{lim}' = {\cal F}_{lim} {\cal F}^{-\alpha}$, with 
$\alpha = 0.23$.  The diagonal line indicates the selection criterion
${\cal F} > {\cal F}_{lim}'$.
}
\label{fig:fluvsflulim}
\end{figure}

The bivariate
distribution of ${\cal F}$ and ${\cal F}_{lim}$ is shown in the
top panel of Figure
\ref{fig:fluvsflulim}.  Obviously
because of the data truncation due to the variation 
in ${\cal F}_{lim}$, simply binning the fluences
to get a distribution will result in a biased distribution.
As explained by Petrosian (1993), a nonparametric method exists to obtain
a single variable distribution from a truncated bivariate distribution.
This method amounts to
using information from untruncated regions to estimate the 
data that was missed due to the truncation.  Clearly, this is only
possible provided the variables are uncorrelated.  The first thing
we must do is test the data for a
correlation between ${\cal F}$ and ${\cal F}_{lim}$.
Using the correlation test designed for use on truncated data
(Efron \& Petrosian 1992), we find that the probability that the
data are uncorrelated is $2.3\times 10^{-5}$.  As discussed in
\S \ref{sec-interp}, since ${\cal F}_{lim}$
is approximately proportional to the duration, 
a correlation test
involving ${\cal F}$ and ${\cal F}_{lim}$ effectively tests the
correlation between fluence and duration.  The results indicate that
the fluence and the duration are positively correlated with each other.
If the fluence is a good measure of distance,
this result seems to be in the opposite sense of that expected
from cosmological time dilation 
(c.f. Norris 1995).
However, there are a number of factors which could complicate this
interpretation (see Lee \& Petrosian 1996b).

In order to go further, one must resort to parameterizing the
correlation.  As we have done before 
(Lee, Petrosian, \& McTiernan 1993; 1995; LP),
we use a simple power law parameterization.
Briefly, we transform ${\cal F}_{lim}$ into 
${\cal F}_{lim}' = {\cal F}_{lim} {\cal F}^{-\alpha}$ and vary $\alpha$
until the correlation between ${\cal F}$ and ${\cal F}_{lim}'$
disappears.  This requirement gives a well-defined value for $\alpha$.
We find $\alpha = 0.22\pm0.07$, with  
the error interval indicating the
$\pm90$\% confidence limits on $\alpha$.  The data truncation boundaries are
transformed accordingly.  The resulting bivariate distribution,
which now contains uncorrelated variables,
is shown in the bottom panel of Figure \ref{fig:fluvsflulim}, which
can then be readily integrated over ${\cal F}_{lim}'$ with the methods 
described in LP.
Using this technique, we obtain the cumulative and differential
distributions of ${\cal F}$, 
along with the logarithmic slope of the cumulative distribution as a function of
${\cal F}$ (Fig. \ref{fig:fludist}).   
The qualitative shapes of these distributions persist even if
we use different samples of bursts corresponding to higher 
fluence limits.  We justify the particular parameterization 
chosen here (power law) by noting that similar values of $\alpha$ are found
for data subsets chosen from various ranges in ${\cal F}$, and in
any case the dispersion in the data is such that there would be
little justification for a more complicated
parameterization.

Dividing our best estimate of the differential distribution
$n({\cal F})$ by what would have been obtained without consideration of the
truncation and correlation 
gives the trigger efficiency as a function of fluence,
which is plotted in Figure \ref{fig:trigeff} along with the ratio of
the observed number of bursts $N_{obs}(>{\cal F})$ to the total
number of bursts $N(>{\cal F})$ greater than a given fluence ${\cal F}$.  
Our derived efficiency can be compared to the
results of in't Zand \& Fenimore (1996) and Bloom, Fenimore, \& in't Zand (1996),
who utilize a quite different
approach.  Rather than starting with the data and working 
backwards through the selection effects to derive the distributions,
in't Zand \& Fenimore use Monte Carlo simulations of a sample of
bursts with a distribution of temporal and spectral shapes
as observed by BATSE.  Then assuming a cosmological origin for the
sources, they 
predict the trigger efficiency of BATSE as a function of fluence.
In contrast, our ``efficiency'' is purely
empirical and involves no model assumptions.  
Using these trigger efficiencies,
Bloom et al. (1996) correct the observed fluence distributions to obtain the
true distributions.  
Their results are qualitatively similar to ours, although their
$\log{N}$-$\log{{\cal F}}$ curve has a very steep upturn at low fluences
which is not evidenced in our curves.
This difference may arise as a result of their assumption that
any correlations between burst characteristics are solely a
result of cosmological effects such as time dilation and redshifting.
However, it would be very difficult to
reconcile the strong positive correlation that we found 
between fluence and duration
with any non-evolving cosmological population of bursts.

\section{Discussion and Conclusions} \label{sec-fludiscuss}

\begin{figure}
\psfig{file=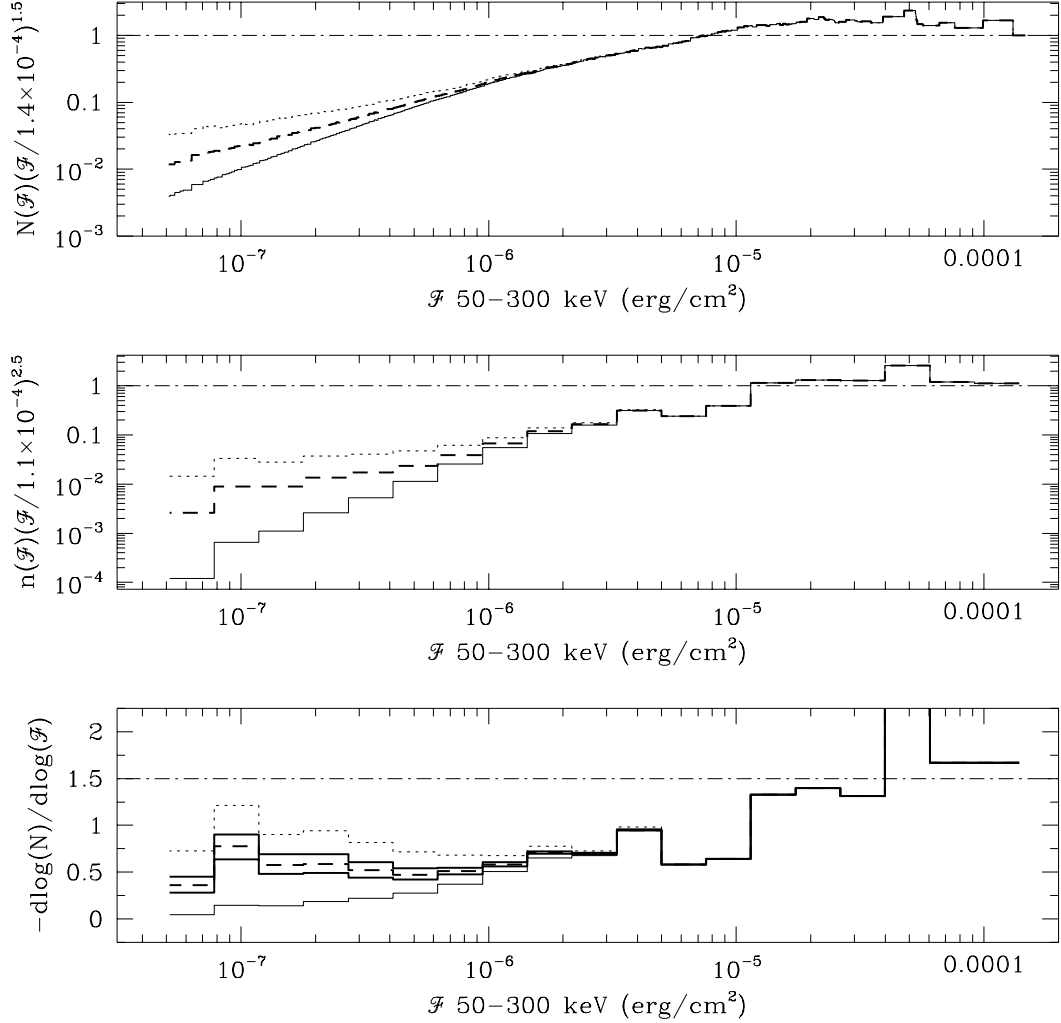,width=5.875in}
\caption{
{\em top panel:} Cumulative distribution $N({\cal F}){\cal F}^{1.5}$ of
fluence, showing the deviation from HISE.  
The solid histogram is the distribution that would have been
obtained without consideration of selection effects or correlations.
The dotted histogram is the distribution that would have been obtained 
by accounting for selection effects but neglecting correlation.
The solid histogram is the distribution obtained when accounting
for both effects.  The dot-dashed line indicates the HISE prediction
of logarithmic slope -1.5. {\em middle panel:} The differential distribution
multiplied by ${\cal F}^{5/2}$, which shows the deviation from the HISE
prediction (dot-dashed line).  {\em bottom panel:} Logarithmic slope of 
$N$ as a function
of ${\cal F}$.  The heavy solid lines show the 90\% confidence limits
on the transformation used to remove the correlation.
}
\label{fig:fludist}
\end{figure}

\begin{figure}
\psfig{file=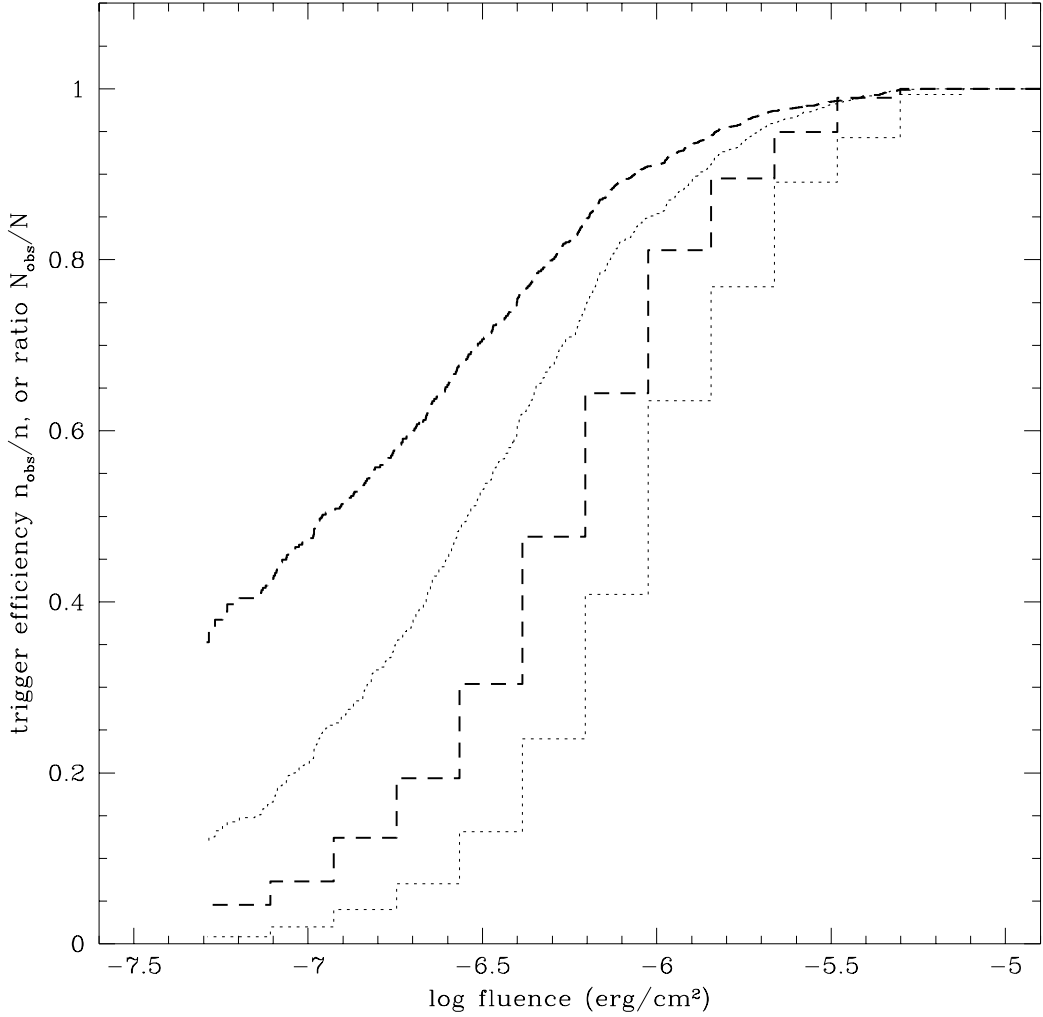,width=5.875in}
\caption{
The lower two histograms show the trigger efficiency $n_{obs}({\cal F})/n({\cal F})$
as a function of fluence,
while the upper two histograms show the ratio of the cumulative number of
bursts observed $N_{obs}(>{\cal F})$ to the true cumulative number of bursts
$N(>{\cal F})$.
The dashed
lines depict our best estimates of the trigger efficiency and ratio of
observed to total number of bursts, accounting for
both data truncation and correlation.  The dotted lines
depict the efficiency or ratio without accounting
for the effects of correlation.
}
\label{fig:trigeff}
\end{figure}

We have described a robust method of accounting for the selection biases
on the detection of BATSE GRBs based on their fluences, independent of duration
or spectral measures.  Using methods described in our previous publications
we have obtained the variation of the cumulative distribution,
differential distribution, and logarithmic slope of the distribution
as a function of fluence.  The results shown in
Figure \ref{fig:fludist}
reveal that the fluence distribution appears to
show a sharp break from slope -3/2 to about -1/2 at 
${\cal F} \approx 10^{-5}$ erg cm$^{-2}$.
The analogous peak flux 
distributions (see Fig. 7 of LP for
an example) show a slightly more gradual transition in slope.
Our interpretation of this result would be that the time-integrated
luminosity has a narrower distribution than the peak luminosity.
Therefore the fluence may be a better indicator of the burst
distance than the peak flux.
The lack of consistency with the time dilation effect would imply
that either the burst sources are not cosmological or 
models more complicated than the simple no-evolution model are
necessary.

Ignoring the inconsistency for the moment,
fits of the fluence distribution to very simple cosmological models (constant comoving density,
no luminosity evolution, energy spectra either power laws 
or as in Band et al. (1993),
density parameter $\Omega=0$ or 1, $H_0 = 75$~km s$^{-1}$ Mpc$^{-1}$)
give sources of total radiant energy
${\cal E} \sim 10^{52}$~ergs or ${\cal E} \sim 10^{51}$~ergs for
$\Omega=0$ and $\Omega=1$ models, respectively. 
Note that uncertainties in the spectral form can be absorbed into
uncertainties in the luminosity evolution, and in any case the
data are not sensitive enough to definitively distinguish among the models.
A difference between these models and those involving 
peak fluxes is that the inferred maximum redshifts are greater
($z_{max} \approx 3$ for the models
involving power law spectra and $z_{max} \approx 5$ for the models
utilizing the Band spectral form),
a result also noted by
Bloom et al. (1996).
However, adding in the evolution necessary for
agreement with the time dilation results would reduce these inferred
maximum redshifts.

\acknowledgements

We acknowledge W. Azzam and G. Pendleton for useful discussions,
and we
thank the anonymous referee for comments and suggestions which led to
an improved paper.
This work was funded by NASA grants NAGW 2290 and NAG-5 2733.

\clearpage


\begin{thebibliography}{}

\bibitem[Band et al. 1993]{BAN93} 
Band, D. et al. 1993, \apj, 413, 281

\bibitem[Bloom et al. 1996]{Blo96}
Bloom, J. S., Fenimore, E. E., \& in't Zand, J. 
1996, proceedings of the 3rd Huntsville
Gamma-Ray Burst Symposium, in press

\bibitem[Efron \& Petrosian 1992]{EP92}
Efron, B. \& Petrosian, V. 1992, \apj, 399, 345

\bibitem[Lee \& Petrosian 1996a]{LP96a} 
Lee, T. T. \& Petrosian, V. 1996a, \apj, in press (LP)

\bibitem[Lee \& Petrosian 1996b]{LP96b} 
Lee, T. T., \& Petrosian, V., 1996b, \apj, in preparation

\bibitem[Lee et al. 1993]{LPM93} 
Lee, T. T., Petrosian, V., \& McTiernan, J. M. 1993, \apj, 412, 401

\bibitem[Lee et al. 1995]{LPM95} 
Lee, T. T., Petrosian, V., \& McTiernan, J. M. 1995, \apj, 448, 915

\bibitem[in't Zand \& Fenimore 1996]{IF96}
in't Zand, J. J. M., \& Fenimore, E. E. 1996, proceedings of the 3rd Huntsville
Gamma-Ray Burst Symposium, in press

\bibitem[Madras \& Fenimore 1996]{MF96}
Madras, C. D., \& Fenimore, E. E. 1996, proceedings of the 3rd Huntsville
Gamma-Ray Burst Symposium, in press

\bibitem[Meegan et al. 1992]{Mee92} 
Meegan, C. A., Fishman, G. J., Wilson, R. B., Paciesas, W. S., Pendleton, G. N.,
 Horack, J. M., Brock, M. N., \& Kouvelioutou, C. 1992, Nature, 355, 143

\bibitem[Meszaros \& Rees 1993]{MR93} 
Meszaros, P. \& Rees, M. J. 1993, \apj, 405, 278

\bibitem[Meszaros \& Rees 1994]{MR94} 
Meszaros, P. \& Rees, M. J. 1994, \apjl, 430, L93

\bibitem[N2]{Nor95} 
Norris, J. P., Bonnell, J. T., Nemiroff, R. J., Scargle, J. D., Kouvelioutou, C.
, Paciesas, W. S., Meegan, C. A., \& Fishman, G. J 1995, \apj, 439, 542 

\bibitem[Petrosian 1993]{Pet93} 
Petrosian, V. 1993, \apjl, 401, L33

\end{thebibliography}
\end{document}